\documentclass[a4paper]{spie}  

\usepackage{amsmath,amsfonts,amssymb}
\usepackage{graphicx}
\usepackage[colorlinks=true, allcolors=blue]{hyperref}
 \pdfoutput=1
\title{Commissioning and First Observations with Wide FastCam at the Telescopio Carlos S\'anchez}

\author[a,b]{Sergio Velasco}
\author[c]{Urtats Etxegarai}
\author[a,b]{Alejandro Oscoz}
\author[a,b]{Roberto L. L\'opez}
\author[a,b]{Marta Puga}
\author[c]{Gaizka Murga}
\author[d]{Antonio P\'erez-Garrido}
\author[a,b]{Enric Pall\'e}
\author[a,b]{Davide Ricci}
\author[a,b]{Ismael Ayuso}
\author[b]{M\'onica Hern\'andez-S\'anchez} 
\author[a,b]{Nicola Truant}

\affil[a]{Instituto de Astrof\'isica de Canarias, c/V\'ia L\'actea s/n, La Laguna, Tenerife E-38205, Spain.}
\affil[b]{Departamento de Astrof\'isica, Universidad de La Laguna, La Laguna, Tenerife E-38205, Spain.}
\affil[c]{IDOM, Avenida Zarandoa 23, Bilbao, Vizcaya E-48015, Spain.}
\affil[d]{Universidad Polit\'ecnica de Cartagena, Campus Muralla del Mar, Cartagena, Murcia E-30202, Spain.  }

\authorinfo{Further author information: \\S.V.: E-mail: svelasco@iac.es}

\begin{document} 
\maketitle

\begin{abstract}
The FastCam instrument platform, jointly developed by the IAC and the UPCT, allows, in real-time, acquisition, selection and storage of images with a resolution that reaches the diffraction limit of medium-sized telescopes. FastCam incorporates a specially designed software package to analyse series of tens of thousands of images in parallel with the data acquisition at the telescope. Wide FastCam is a new instrument that, using the same software for data acquisition, does not look for lucky imaging but fast observations in a much larger field of view. Here we describe the commissioning process and first observations with Wide FastCam at the Telescopio Carlos S\'anchez (TCS) in the Observatorio del Teide.
 
\end{abstract}


\section{INTRODUCTION}
\label{sec:intro}  

Wide-FastCam (WFC) is an adaptation of the FastCam platform (Oscoz et al. \cite[2008]{2008SPIE.7014E.137O}) with the objective of obtaining wide field images ($8'\times8'$ FoV, $0.5''/$px and a $1k\times1k$ Andor Ixon 888 EMCCD) at a high temporal resolution (8 fps) with real time processing capacity. The large field of view guarantees the presence of a photometric calibration star in the majority of the pointings of the telescope. These features make WFC a unique instrument in the world for the observation of transient phenomena: no other instrument can process in real time the tens of thousands of images obtained each night with such a FOV in the optical range. 

An initial prototype, evolution of FastCam but with different optics and detector, was developed at the Instituto de Astrof\'isica de Canarias (IAC) as a test bed of the instrument we are presenting here. As a prototype, WFC was installed at the TCS on several occasions, helping to investigate and solve some opto-mechanical issues. Even in this initial phase, WFC provided quite valuable scientific data.

   \begin{figure} [ht!]
   \begin{center}
   \includegraphics[height=8cm]{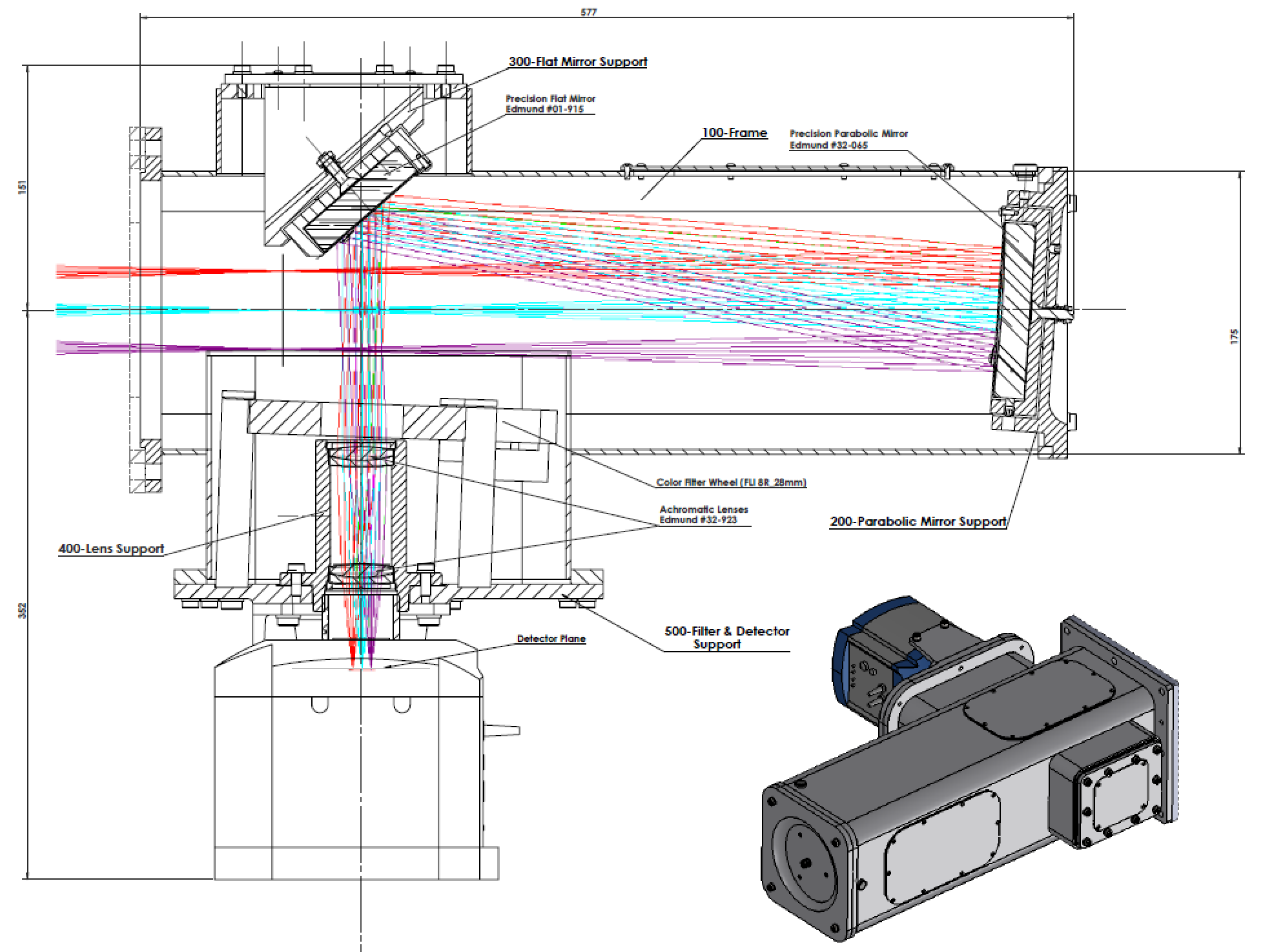}
   \end{center}
   \caption[example] 
   { \label{fig:img1} 
Section view of the instrument with the ray tracing superimposed.}
   \end{figure} 

The final version is a much more compact and compensated instrument with a new optical design (see Murga et al. \cite[2014]{2014SPIE.9147E..6QM}). Manufactured to be installed at the Cassegrain focus of the TCS, its high versatility allows its installation as a visiting instrument in many other  telescopes of the Observatorios de Canarias. WFC shares the same control system than FastCam, developed by the IAC and the UPCT (Cartagena, Spain) and unique in its capabilities among this kind of instruments, providing tens of thousands of images per night which can be analyzed in real time. Moreover, some semi-automated pipelines have been developed for the analysis of differential photometry in the data cubes being able to produce and offer via internet light curves in real time.

With the remotization of the TCS telescope, WFC was one of the first instruments at this telescope to offer to the astronomers the possibility of performing the observations through an internet connection from any point in the world. 

\section{INSTRUMENT}

The main component of Wide FastCam is the $1k\times1k$ Andor’s 888 EMCCD, with an E2V sensor reaching an 8 fps readout when used without binning or windowing regions. This detector has a pixel size of $13.3\,\mu$m meaning a total size of $13.6\,$mm, and hence a focal reduction of at least 3.6 is needed to render the desired $8\times8$ arc minutes field of view.

\begin{figure} [ht!]
   \begin{center}
   \includegraphics[height=7cm]{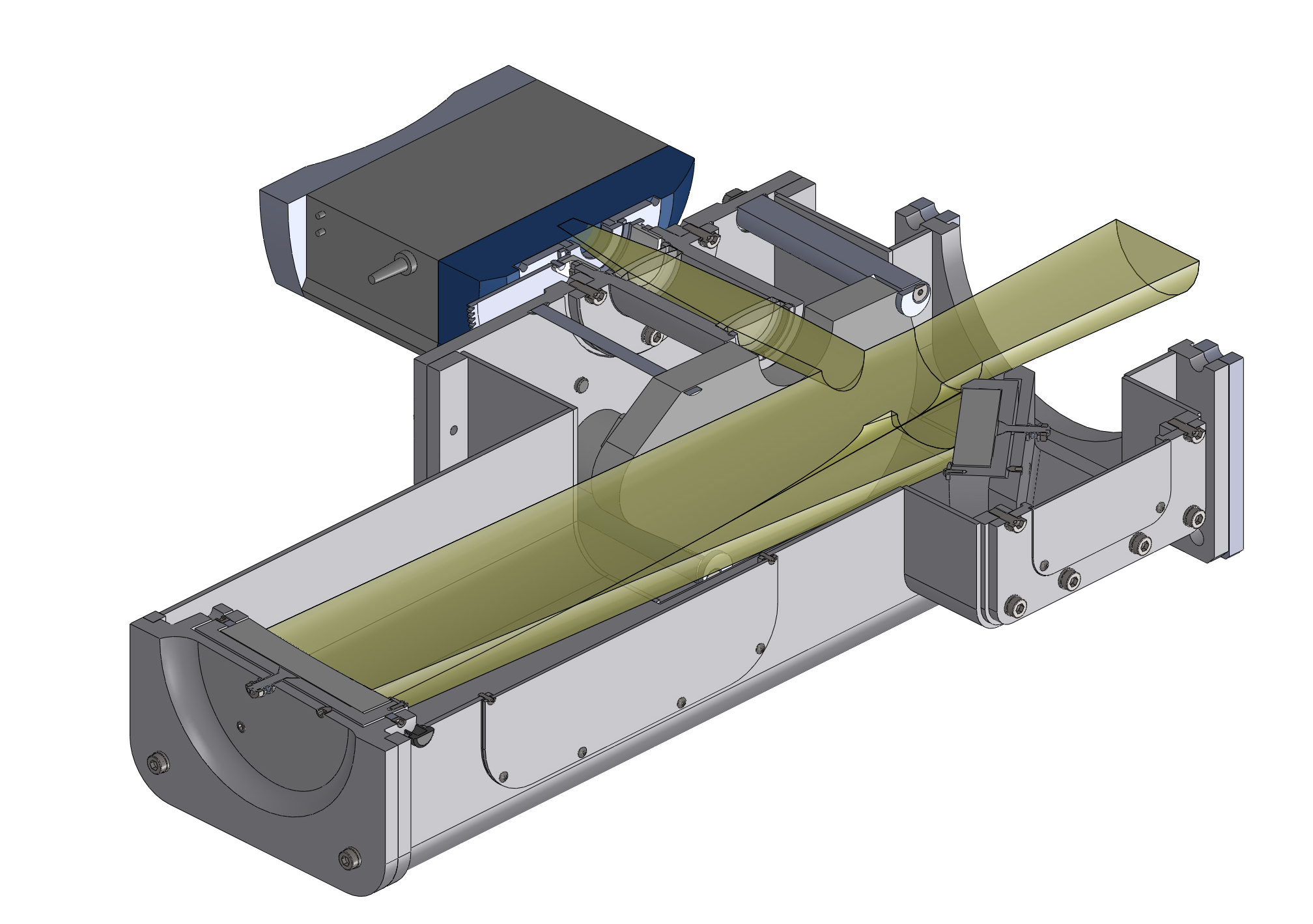}
   \end{center}
   \caption[example] 
   { \label{fig:img2} 
Section view of the instrument's 3D model.}
   \end{figure}

The focal reduction is achieved by a classical collimator-camera configuration, see figures \ref{fig:img1} and \ref{fig:img2}. Due to space limitations in the distance below the telescope’s Cassegrain focus the encapsulated design had to be done with some length restrictions.

The optical components can be aligned in a very simple way thanks to the positioning screws located out of the structure. A filter wheel has been integrated within the instrument.

\section{COMMISSIONING PLAN}

The scientific validation and commissioning tests were focused on characterizing the behaviour of the EMCCD detector, studying its gain linearity, the dynamic and photometric ranges and its sensitivity besides of making several astrometric tests.

Since the develop of FastCam, the high optical spatial resolution group at the IAC has worked with EMCCDs, developing new instruments, such as AOLI (see Velasco et al. \cite[2015]{2015hsa8.conf..850V} and Mackay et al. \cite[2014]{2014SPIE.9147E..1TM}), and producing many very valuable science results (e.g. Velasco et al. \cite[2016]{2016MNRAS.tmp..855V}, Femen\'ia et al. \cite[2011]{2011MNRAS.413.1524F} and Labadie et al. \cite[2011]{2011A&A...526A.144L}). As a needed complement of these observations, along the years a huge amount of laboratory and telescope hours have been devoted to characterize the EMCCD detectors in use by these instruments. Once this know-how was settled, the natural step was to bring up a reliable method to calibrate and characterize them. WFC's commissioning followed this method, being the first EMCCD instrument in doing it and setting a precedent for other present and future instruments. The fundamentals of the method are:

\begin{itemize}
\item The high availability of on-sky calibration nights at the TCS, allowing a fully detailed commissioning with the equivalent of tens of night hours. The strength of our method comes from the use of natural stars with different atmospheric conditions, rather than calibrated light sources at the lab, becoming the EMCCD's commissioning of WFC unique.
\item The gain linearity was measured with some standard stars with known and calibrated magnitudes in the optical bands. We selected stars within a range of magnitudes from $0th$ to $14th$ in the \textit{I} and \textit{V} bands and observed them varying the EM gain of the detector and along many different zenith altitudes. The results are compared with dome flats obtained with a similar procedure.
\item The dynamic and photometric range are studied by selecting five stellar populated regions at different declination positions and visible during the whole night.
\item The astrometric and sensitivity tests were performed with some selected stellar crowded regions.
\end{itemize}

   \begin{figure} [ht!]
   \begin{center}
   \begin{tabular}{c}
   \includegraphics[height=7cm]{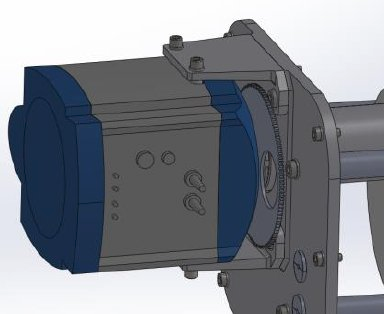}
   \includegraphics[height=5cm]{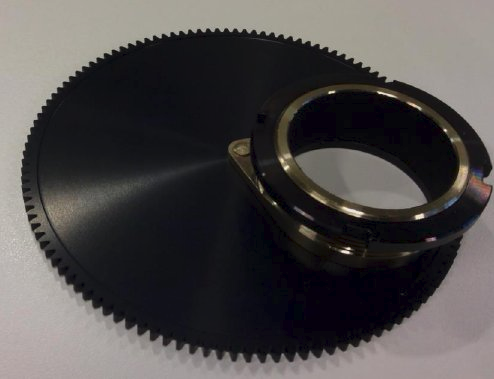}
   \end{tabular}
   \end{center}
   \caption[example] 
   { \label{fig:img3} 
Scheme of the external shutter position at Wide FastCam (left) and external shutter finally designed for the instrument (right).}
   \end{figure}

  The Andor Ixon 888 lacks mechanical shutter and this affects the precision of the bias. To solve it, an external physical bias mechanism, see figure \ref{fig:img3}, was designed and tested during the commissioning.  

\section{FIRST LIGHT}

\begin{figure} [ht!]
   \begin{center}
   \includegraphics[height=7cm]{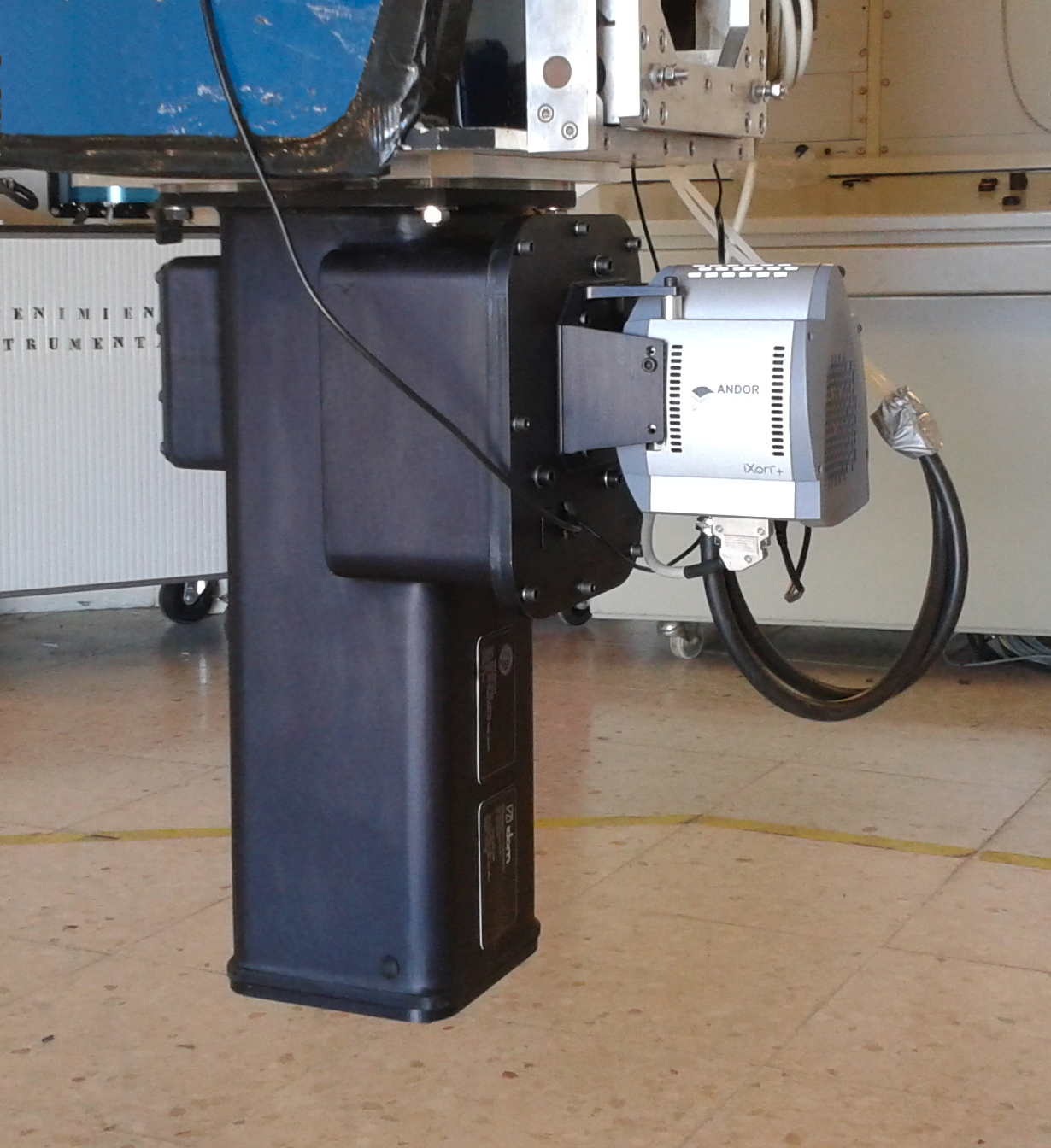}
   \end{center}
   \caption[example] 
   { \label{fig:img5} 
Wide FastCam mounted at the TCS.}
   \end{figure}

The final encapsulated and compact concept of Wide FastCam saw first light at TCS in August 2014, as shown in figure \ref{fig:img5}.

\begin{figure} [ht]
   \begin{center}
   \begin{tabular}{c}
    \includegraphics[height=8cm]{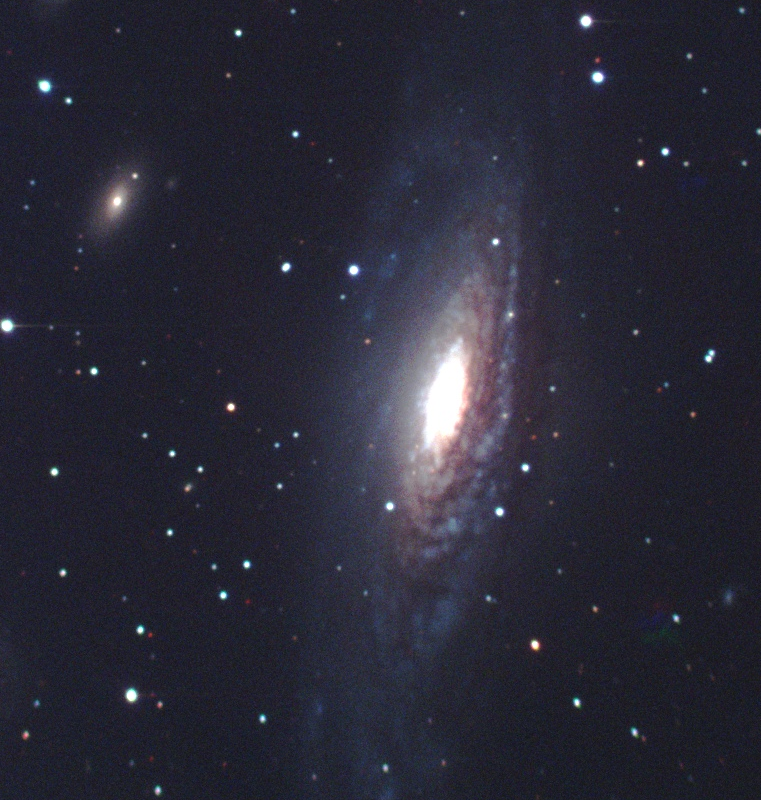}
    \includegraphics[height=8cm]{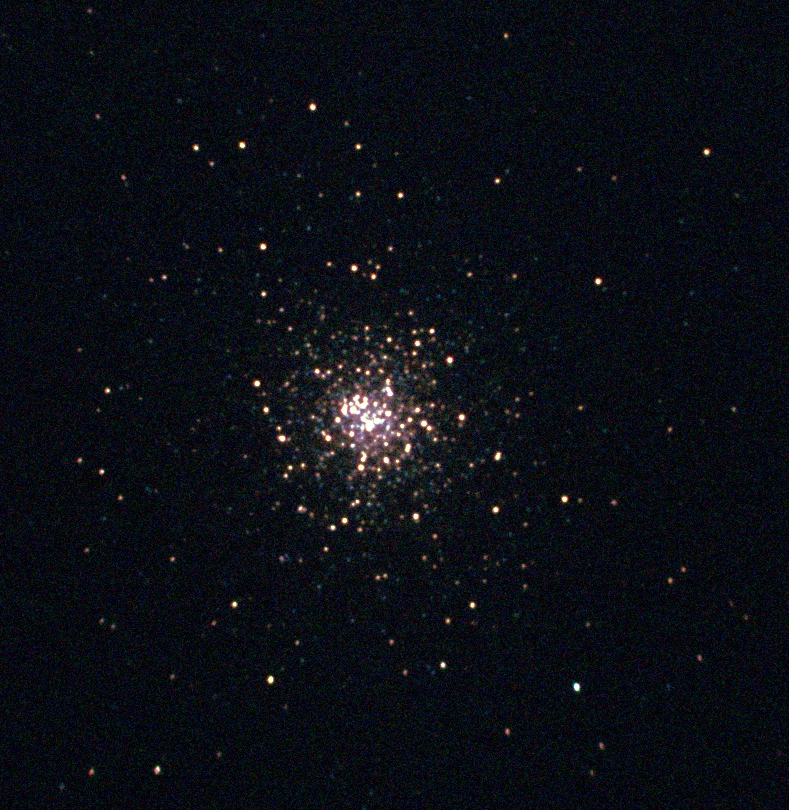}
   \end{tabular}
     \end{center}
   \caption[example] 
   { \label{fig:img6} 
RGB composed images of NGC7331(left) and M13 (right).}
   \end{figure}

The TCS is a $1.5\,$m telescope initially designed for small FoV IR observations. Regardless of that, the TCS+WFC combination has an optical quality good enough to deliver superb images in the optical across the whole FoV (see figure \ref{fig:img6}).

Since its commissioning, WFC has been regularly used not only for transient observations but also for several other research fields.  

\section{SOME RESULTS}

The results obtained from the Wide FastCam commissioning prove the high potential and versatility of this instrument, and the validity of its scientific observations.

\begin{figure} [ht]
   \begin{center}
    \begin{tabular}{c}
    \includegraphics[height=5cm]{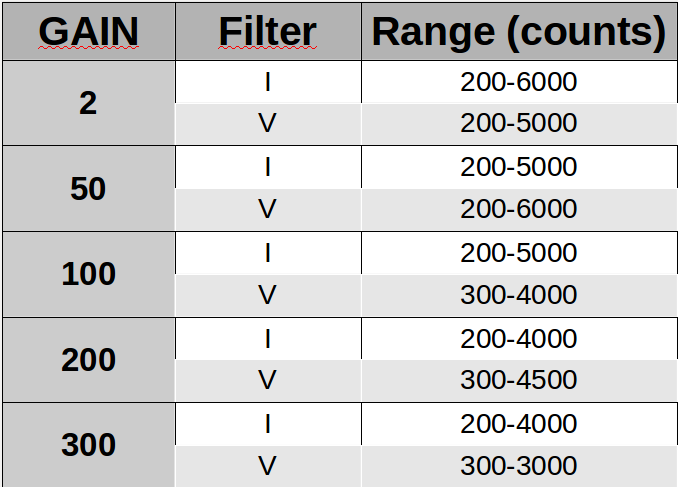}
    \includegraphics[height=5cm]{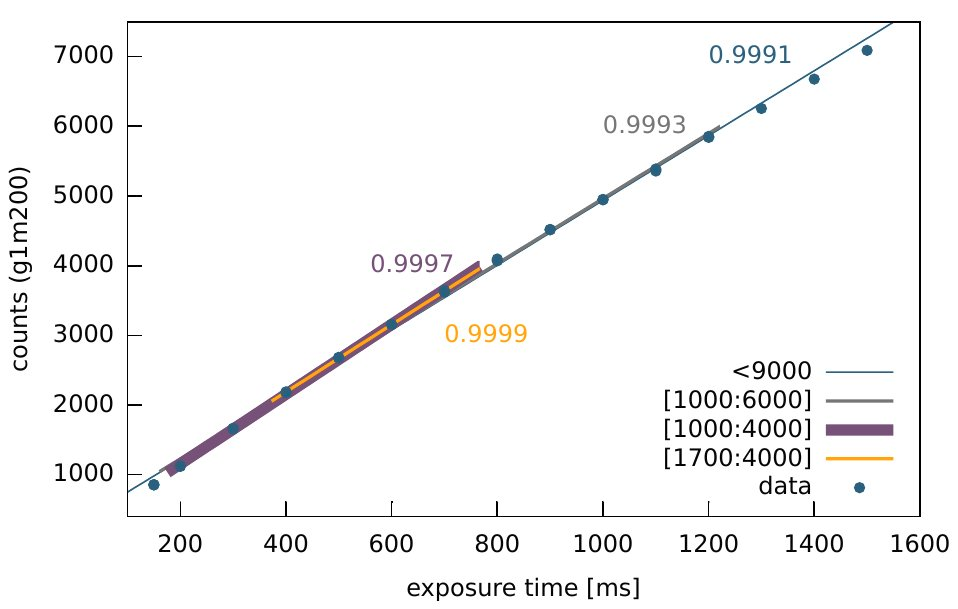}
   \end{tabular}
     \end{center}
   \caption[example] 
   { \label{fig:img8} 
Photometric linear range evaluated for I and V filters and different EM gains and linearity  for gain 200 (right).}
   \end{figure}

\begin{figure} [ht]
	\begin{center}
    \begin{tabular}{c}
    \includegraphics[height=8cm]{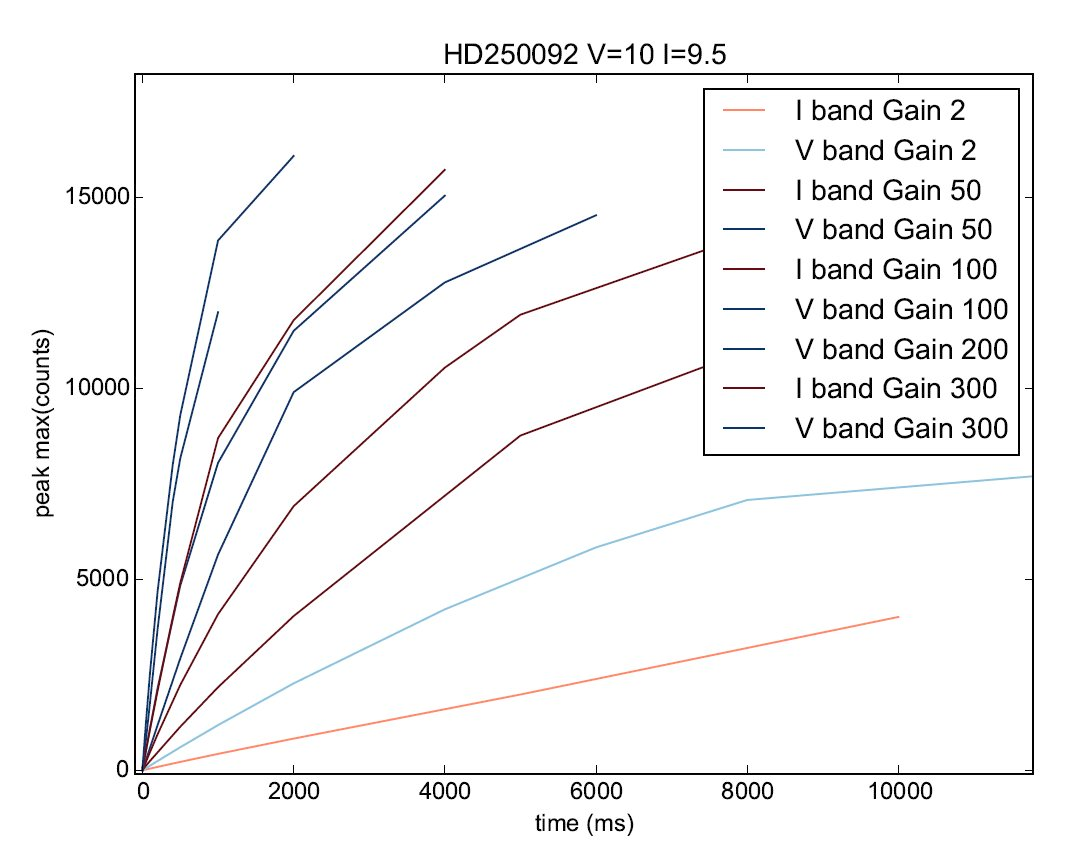}
       \end{tabular}
	\end{center}
	\caption[example] 
	{ \label{fig:img9} 
		On-sky linearity and sensitivity test.}
\end{figure}

Among other interesting results, the linearity of the WFC's EMCCD detector has been found to be dependent on gain and exposure time, as seen in figures \ref{fig:img8} and \ref{fig:img9}. 

\begin{figure} [ht]
	\begin{center}
		\includegraphics[height=8cm]{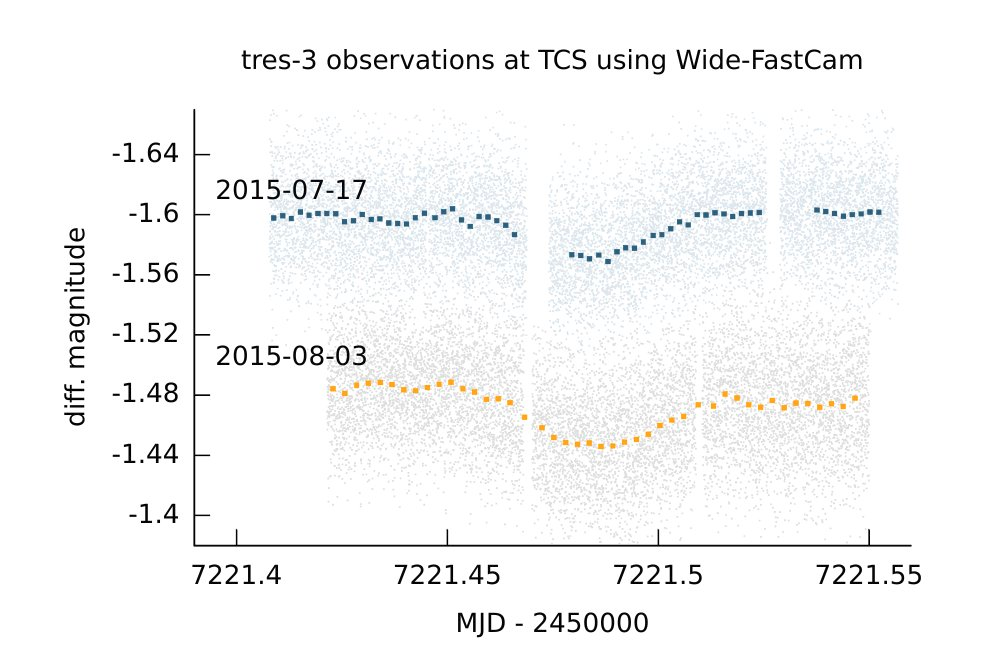}
	\end{center}
	\caption[example] 
	{ \label{fig:img11} 
		Exoplanet TrES-3 transit light-curve with WFC data.}
\end{figure}

As a brief example, we present in figure \ref{fig:img11} the TrES-3 exoplanet light curve. The superb quality of WFC is demonstrated with the extreme time-resolution of the data collected with a quite reduced noise deviation.

\section{CONCLUSIONS AND FURTHER WORK}

As it has been tested, Wide FastCam offers a huge branch of possibilities on high time resolution astronomy. The high sensitivity of the EMCCD camera joint to a wide FoV provide the scientific community with a powerful tool to understand the behaviour and nature of transients astrophysical phenomena.

Our next goal is to provide the software with a guiding system corrector extracting the shifts information form the science images, allowing us to reduce the photometrical noise introduced by the displacement of the sources.

The modular concept used in Wide FastCam allows an easy adaptation to other telescopes like NOT, WHT or GTC, were FastCam has already been successfully tested. This versatility allows Wide FastCam  to perform large programs of exoplanet searching and fast wide field differential photometry.

\acknowledgments 
This article is based on observations made with the Telescopio Carlos S\'anchez operated on the island of Tenerife by the Instituto de Astrof\'isica de Canarias in the Spanish Observatorio del Teide.
\bibliography{report} 
\bibliographystyle{spiebib} 

\end{document}